\documentclass{aastex}
\usepackage{spr-astr-addons}
\usepackage{url}
\RequirePackage{color}

\usepackage{bm}% bold math

\def\be{\begin{equation}}
 \def\ee{\end{equation}}
\def\bea{\begin{eqnarray}}
\def\eea{\end{eqnarray}}

\def\l{\label}
\def\o{\over}
\def\om{\omega}
\def\Om{\Omega}
\def\p{\phi}

\def\R{\rho}
\def\na{\nabla}
\def\L{\Lambda}
\def\s{\sigma}

\begin{document}

\title{ On the  holographic dark energy  in  chameleon scalar-tensor cosmology}
\slugcomment{Not to appear in Nonlearned J., 45.}
%% Running heads
\shorttitle{Short article title}

\shortauthors{Saaidi and Sheikhahmadi}

\author{Kh. Saaidi$^{\dag}$\altaffilmark{1}} \and
\author{H. Sheikhahmadi$^{\dag \maltese}$ \altaffilmark{2}} \and
\author{T. Golanbari$^{\dag}$ \altaffilmark{3}} \and
\author{S. W. Rabiei$^{\dag}$ \altaffilmark{4}}
\affil{$^{\dag}$Department of Physics, Faculty of Science, University of Kurdistan, Sanandaj, Iran.} \and
\affil{$^{\maltese}$Young Researchers and Elites Club, Sanandaj Branch, Islamic Azad University, Sanandaj, Iran.}

%\email{\emaila}

\altaffiltext{1}{ksaaidi@uok.ac.ir}
\altaffiltext{2}{h.sh.ahmadi@uok.ac.ir or @gmail.com}
\altaffiltext{3}{t.golanbari@uok.ac.ir or @gmail.com}
\altaffiltext{4}{w.rabiei@gmail.com}

\begin{abstract}
We study the holographic dark energy (HDE) model in generalized Brans-Dicke scenario with a non-minimal coupling between the scalar field and matter lagrangian namely Chameleon Brans Dicke (CBD) mechanism. In this study we consider the  interacting and non-interacting cases   for  two different cutoffs. The physical quantities of the  model such as, equation of state (EoS) parameter, deceleration parameter and the  evolution equation of dimensionless parameter of dark energy are obtained. We shall show that this model can describe the dynamical evolution of fraction parameter of dark energy in all epochs. Also we find  the  EoS parameter can cross the phantom divide line by suitable choices of parameters without any mines kinetic energy  term.
\end{abstract}

%\section*{}
%\label{sec:intro}

{\section{Introduction}}
Cosmological  and astrophysical observational data risen from supernovae type Ia (SNIa){\citep{2, 2a, 2b, 2c}}, Cosmic Microwave Background Radiation (CMBR) {\citep{3}} and Sloan Digital Sky Survey (SDSS) {\citep{4, 4a, 4b, 4c, 4d}} indicate that the Universe is in  accelerated expansion regime.\\
 There are two approaches to justify the source of accelerating phase of the Universe. Some people look for the source of this acceleration in the geometrical part of the Hilbert-Einstein action and have studied the modified gravity {\citep{f, f1, f2, f3, f4, f5, f7, f6}}. As a second way, some researchers propose an eccentric form of matter namely dark energy (DE) {\citep{5, 5a, 5b, 5c, 5d, 5e}}. Although the nature and origin of the DE are ambiguous for researchers up to now, but people proposed some useful candidates which could satisfy both theoretical and observational results {\citep{6, 6a, 6b, 7, 7a, 7b}}. Amongst these proposals cosmological constant model ,$\Lambda$, is the fundamental block. It is clear that this model suffers from two well known problems i.e. the "cosmological coincidence problem"   and "the fine tuning problem", we refer the reader for more details to {\citep{8, 8a, 8b, 9, 6a, 10}}.

 Recently scalar field models attract more attentions to investigate the behavior and nature of the DE. Most of DE models treat scalar fields as DE component with a dynamical equation of state.
The basic  dynamical DE proposal  which is called "quintessence" model consider the slow-roll down of a scalar field and suggests  an energy form with negative pressure {\citep{11, 11b, 11c, 12, 13, 14, 15, 16, 17}}.
Another suitable framework to investigate the behavior of DE is
chameleon mechanism {\citep{20, 20a, 20b}}. In this mechanism scalar
field has a non-minimal coupling with  matter sector. Chameleon mechanism
provides an alternative mechanism to satisfy the constraints from local test of gravity. For more acquaintance we refer the reader to {\citep{21, 21a, 21b, 22}} and the works which are there.

In recent years some problems risen with BD model or chameleon model alone in cosmology, so some researchers such as, {\citep{23}} and {\citep{24}}, have studied the frame work which scalar field has non-minimal coupling with both the geometry and matter sectors so-called CBD mechanism. For future studies about CBD mechanism we refer the reader to {\citep{25, 25a, 25b, 25c, 26, 26a, 26b}}.

Also some other  DE models ( which arise from space-time fluctuations and quantum gravity) such as, agegraphic DE models (original and new model) {\citep{18, 18a}}, and holographic DE models (HDE) {\citep{19, 19a, 19b, 19c, 19d}} have been introduced.
Amongst various scenarios to describe the accelerating expansion of the Universe, the holographic dark energy (HDE) model have got more attentions. Based on the quantum gravity, the density of HDE is regarded as zero-point energy density and defined versus $L$, the size of the Universe, as follows {\citep{27}}
\begin{equation} \label{1'}
\rho_{\Lambda}= 3c^2 M_p^2L^{-2},
\end{equation}
where $L$ is the size of the Universe,  $3c^2$ is  introduced for convenience and $M_p=1/\sqrt{8\pi G}$ is  the reduced Planck mass where $G$ is the Newtonian gravitational constant. The HDE is studied for different choices of  infrared cutoffs of the Universe, e.g., Hubble horizon, particle horizon, future event horizon {\citep{28, 28a, 28b, 28c}}. In the context of standard model of cosmology and for  a non interacting case, if we take the particle horizon as an infrared cutoff, the accelerated expansion of the  Universe cannot be explained {\citep{29}}, and if the Hubble horizon chooses as the cutoff, then an appropriate equation of state parameter for dark energy cannot be derived {\citep{30}}, only the future event horizon has reasonable behavior which is done in {\citep{29, 31}.

     Besides of these studies, some researchers studied the HDE model in  CBD scenario  for future event horizon   and  didn't recognize the basic feature of this model \footnote{Some features of this model is studied in Phys. Lett. B {\bf 697}, 285 (2011). But the fundamental property of the model is recognized wrong and then the whole results of it  are incorrect. We tried to correct the mentioned paper and extended our investigation to other features of the model}. These incorrect studies and existence another infrared cutoff, conformal-age-like cutoff,  motivated us to investigate the HDE in generalized CBD model of cosmology.  In this paper we consider the interacting and non interacting cases of HDE model in generalized CBD scenario for two different cutoffs, future event horizon and conformal-age-like cutoff respectively.

The scheme of this paper is as follows:   In Sec.\;2  HDE model in  generalized CBD scenario is considered. The scalar  and gravitational field equations are obtained and also conservation equation of energy density is modified.
In Sec.\;3 Future event horizon has been considered  as an IR cutoff and therefore the EoS, deceleration parameters and other relevant quantities for both interacting and noninteracting cases have obtained. In Sec.\;4 A new cutoff so called  "{\it conformal-age-like length}" cut off, which arises from  the four dimensional space-time volume at cosmic time $t$ in the flat Fridmann, Limature, Robertson, Walker (FLRW) Universe is considered. The last section is devoted to some concluding remarks and discussions.\\
%=================================================================
%==============section 2 ( Non-interacting HDE in genaralized CBD Cosmology)=====
%=================================================================

{\section{Field equations and conservation relation of energy }}

 Generalized CBD theory in which scalar field
has  non-minimal coupling with both  geometry and matter sectors is considered as
\bea\label{1}
A&=&\int d^4x\sqrt{-g}\bigg[ {1\o 2}\Big\{\phi R-  \nonumber
\frac{\omega(\phi)}{\phi}\na_{a}\phi\na^{a}\phi-V(\phi)\Big \}\\
 &+& f(\phi){L}_{m}\bigg],
\eea
where R is the Ricci  scalar,  $\omega(\phi)$ is the  CBD parameter ( i.e., as a coupling function), and
${L}_{m}$ is the lagrangian of the matter. $\phi$ is CBD scalar field and $V(\phi)$ is inverse  power law potential which defined as  $V(\phi)=M^{4+\nu}/\phi^{\nu}$ with a positive constant ,$\nu$,  {\citep{11, 32, 33, 34}}.
Note that the last term in the action indicates the interaction
between the matter and some arbitrary function $f(\phi)$
of the CBD scalar field. One can obtain the gravitational and scalar  field equations of motion  by
varying  the action (\ref{1}) with respect to (w.r.t)  $g^{ab}$  and $\phi$ respectively. The gravitational field equation is
\begin{equation}\l{2}
\phi G_{ab}\equiv  \phi\Big[R_{ab} - \frac{1}{2}g_{ab}R\Big]=f(\phi)T_{ab}+ T^{\phi}_{ab},
\end{equation}
where $G_{ab}$ is the  Einstein tensor, $R_{ab}$ is the   Ricci tensor, $T_{ab}$ is the
energy-momentum tensor of the matter which is given by
\begin{equation}\label{3}
T_{ab} = -{2\over \sqrt{-g}}{\delta[\sqrt{-g}L_m ] \over \delta g^{ab}},
\end{equation}
 and $T^{\phi}_{ab}$ is defined as
\bea\l{4}
T^{\phi}_{ab}&=&\frac{\omega}{\phi} \Big[ \nabla_a \phi \nabla_b \phi -\frac{1}{2} g_{ab}\nabla_c \phi \nabla^c \phi \nonumber
 \Big]+\nabla_a \nabla_b \phi\\
 &-& g_{ab}\nabla_a \nabla^a  \phi- \frac{1}{2}g_{ab}V(\phi).
\eea
We suppose that all components of matter (cold dark mater and DE) are perfect fluid and then we can introduce  the  stress-energy tensor  of matter as
\begin{equation}\label{5}
T_{ab}=(\rho_t+p_t)u_{a}u_{b}+p_tg_{ab},
\end{equation}
where $u^{\mu}$ is the four-vector velocity of the fluids and  $\rho_{t}$ and $p_{t}$
indicate the total energy density and  pressure respectively.

 The scalar field equation of motion  is obtained as
\bea\l{6}\nonumber
\Big[2\om(\p)+3\Big]\nabla_a \nabla^a \phi &=& f(\phi)T-2\p f'(\phi) L_m + \phi V'(\phi) \\
 &-& 2V(\phi)-\om'(\p) \nabla_a\p\nabla^a \phi,
\eea
where $T$ is the trace of $T_{ab}$ and prime denotes derivative with respect to $\p$. Setting  $f(\phi)=1$ and $V(\phi) =0$, the above equations reduce to those  of Ref. {\citep{25, 25a, 25b, 25c}}.

It is seen that for solving (\ref{6}) we need an explicit form of matter Lagrangian, $ L_m$.
  The Bianchi identities, together with the identity
$(\square\nabla_a - \nabla_a\square  ) V_c = R_{ab}\nabla^bV_c$, imply the non-(covariant)
conservation law
\be\label{7}
\nabla^aT_{ab}=\big[T^a_b-\delta^a_b  L_m  \big]{\nabla_a \ln(f)},
\ee
and, as expected, in the limit $f(\p)$ = constant, one recovers the
conservation law $\nabla_aT^{ab} = 0$  \footnote{Eq.\;(\ref{7})  is not recognized correctly  in \cite{1}, therefore  the results which obtained in their work is not correct.}.

We consider, the homogeneous and isotropic FLRW background
metric with line element
\begin{equation}\label{8}
ds^{2}=-dt^{2}+a^{2}(t)\left[\frac{dr^{2}}{1-kr^{2}}+r^{2}(d\theta^{2}+\sin^{2}\theta
d\phi^{2})\right].
\end{equation}
Where $a(t)$ is the scale factor and $k=-1, 0, +1$ indicate the open,
flat and close Universe respectively. From Eqs.\;(\ref{2}), (\ref{5}) and (\ref{8}), one can obtain the components of gravitational equation as
\begin{equation}\label{9}
3\left(H^2+\frac{k}{a^{2}}\right)=\frac{f(\phi)}{\phi}\rho_{t}-
3\frac{\dot{a}}{a}(\frac{\dot{\phi}}{\phi})
+\frac{\omega(\phi)}{2}\frac{\dot{\phi}^{2}}{\phi^{2}}+\frac{V(\phi)}{2\phi},
\end{equation}

\begin{equation}\label{10}
2\dot{H}+ 3H^2+\frac{k}{a^{2}}=-\frac{f(\phi)}{\phi}p_{t}
-\frac{\omega(\phi)}{2}\frac{\dot{\phi}^{2}}{\phi^{2}}-2\frac{\dot{a}}{a}(\frac{\dot{\phi}}{\phi})
-\frac{\ddot{\phi}}{\phi}+\frac{V(\phi)}{2\phi}.
\end{equation}
Here $H={\dot{a}}/{a}$ is the Hubble parameter, and dot indicates
differentiative w.r.t the cosmic time, $t$.

Whereas perfect fluid is an averaged properties of matter then it is  not necessary to know an exact description of matter, therefore  it is more common to work directly with energy-momentum tensor instead of Lagrangian. But in present model the Lagrangian, $L_m$, is explicitly appeared  in equation of motion of scalar field, (\ref{6}), and we have to know what is it?. It was considered that Eq.\;(\ref{3}) can give us a stress-energy tensor, (\ref{5}), for a perfect fluid with a matter Lagrangian as $L_m=p_t$ {\citep{36, 37, 38, Gh}}, where $p_t$ is the pressure of the fluid. In fact the on-shell action, which is the proper volume integral of the pressure
     \be\l{11}
     A({\rm on-shell}) =  \int\; d^4x \;\sqrt{-g}\; p_t(\mu, s),
     \ee
     give the stress-energy tensor, (\ref{3}), by varying it w.r.t $g^{ab}$.  By adding some surface integral to the above action, the action will change its on-shell value
     without affecting the equation of motion. By considering this fact it is shown that,  Lagrangian is not unique.
     It is found that the other choices can be
 $ -\R_t(n,s)$ and $-na_t(n, {\cal T})$ {\citep{36, 37, 38}} where $\R_t$ is the total density energy, $n$ is the density of particles,    $a_t(n, {\cal T}) = \R_t/n -{\cal T}s$, is the physical free energy, ${\cal T}$ is temperature of fluid and $s$ is entropy. For a complete review see {\citep{38, 39}}.\footnote{ Some  authors have  chosen $ L_m= T/4$ ( here $T$ is the trace of matter  stress-energy tensor), and  eliminate the Lagrangian, $L_m$, in Eq.\;(\ref{6})  According to our definition, this choice  is
 \be\l{12}
 L_m = -{1\o 4}\R_t +{3\o 4}p_t.
\ee
But for the case which matter (perfect fluid) has an interaction with other components of the model, the degeneracy is broken {\citep{Gh}}. In fact it is shown that the degeneracy definition of $\rho_{t}$ and $P_{t}$ are not equivalent, therefore the motion of perfect fluid is geodesic only for lagrangian density $P_{t}$.}
Based on the earlier discussion, Eq.\;(\ref{12}) with together  (\ref{3}) give a   stress-energy tensor of matter  as (\ref{5}).

 In this work, we assume
 \be\l{13}
  L_m = p_t,
 \ee
   so Eq.\;(\ref{6}) is reduced to

\begin{eqnarray}\label{14}\nonumber
\big[2\om(\p)+3\big] \Upsilon &=& \\
\big[\rho_t-3p_t\big]f(\p)&-&\frac{1}{2}\phi f'(\p)p_t +2V(\p)-\phi V'(\p).
\end{eqnarray}
Where $\Upsilon=\Big[\ddot{\phi}+3H\dot{\phi}+{\om'(\p) \o 2\om(\p) +3}\dot{\phi}^{2} \Big]$.
By using  Eqs.\;(\ref{7}) and (\ref{13})  one can attain  conservation equation as
\begin{equation}\label{15}
\dot{\rho_{t}}+3H\rho_{t}(1+\omega_{t})=
-\frac{\dot{f}}{f}(1+\omega_{t})\rho_{t},
\end{equation}
where $\rho_{t}=\rho_{\Lambda}+\rho_{m}$ and  we have used $p_{t}=\omega_{t}\rho_{t}$. Notice $p_{t}=p_{\Lambda}+ p_m$ yields  $p_{t}=\omega_{t}\rho_{t}=\omega_{m}\rho_{m}+\omega_{\Lambda}\rho_{\Lambda}$. So one can rewrite (\ref{15}) as follows
\bea\label{16}
\dot{\rho}_\Lambda+3H\rho_\Lambda\big(1+\omega_{\Lambda}\big)&=&
-\frac{\dot{f}}{f}(1+\omega_{\Lambda})\rho_{\Lambda},\\
 \dot{\rho}_m + 3H \rho_m\big(1+\om_m\big)&=& -\frac{\dot{f}}{f}\big(1+\om_m\big)\rho_{m}.\l{17}
\eea

So according to the original definition of HDE density, Eq.\;(\ref{1'}),  the HDE density in the CBD scenario is defined  as
\be\l{18}
\rho_{\Lambda}=\frac{3c^2 M_{P}^{2}\phi}{L^2},
\ee
 moreover,  critical energy density, $\R_c$, and energy density of curvature, $\R_k$, in the generalized CBD model, are
\bea
\rho_{c}&=&{3\phi H^2},\l{19}\\
\rho_{k}&=&-\frac{3\phi k}{a^2}.\label{20}
\eea

For more convenience we consider  $M_{P}^{2}=1$.
 Therefore energy density parameters are obtained as
\bea\label{21}
\Omega_\Lambda &=&\frac{\rho_\Lambda}{\rho_c}=\frac{c^2}{H^2 L^2},\\
\Omega_k&=&\frac{\rho_k}{\rho_c}=\frac{k}{a^2 H^2},\\\l{22}
\Om_m &=& {\R_m \o \R_c} = {\R_m\o 3H^2\p}.\l{23}
\eea
Based on these dimensionless density parameters, we can rewrite Eq.\;(\ref{9}) as
\be\l{24}
f(\p)\Om_t + \Om_k+{1\o 2}\Om_V + \Om_{\p} = 1,
\ee
where
\bea\l{25}
\Om_{\p}&=&{1\o 3H^2}\bigg[\frac{\omega(\phi)}{2}\frac{\dot{\phi}^{2}}{\phi^{2}} -
3\frac{\dot{a}}{a}(\frac{\dot{\phi}}{\phi})\bigg],\\
\Om_V&=& {V(\p) \o 3H^2\p}.\l{26}
\eea

\section{Future event horizon as an IR cut off}
 In this Section we want to calculate the physical quantities for a special cut off namely future event horizon.
Event horizon  is  defined as $L=a r(t)$, where $a$ is scale factor and $r(t)$ is
\begin{equation}\label{27}
r(t)=\frac{1}{\sqrt{|k|}}{\rm sinn}(\sqrt{|k|}y) = \left\{ \begin{array}{l}
{\rm sin}(y)\,\,\,;~k =  + 1 \\
y\,\,\,\,\,\,\,\,\,\,\,\,\,\,;~~k=0  \\
{\rm sinh}(y)\,\,\,;~k= -1. \\
\end{array} \right.
\end{equation}
In this relation $y={R_h}/a(t)$ where $R_h$ is future event horizon and ${\rm sinn}(\sqrt{|k|}y)$ indicates elliptic functions.
So taking derivative $L$ w.r.t the cosmic time  and using  Eq.\;(\ref{27}) yeilds
\begin{equation}\label{28}
\dot{L}=HL+a\dot{r}(t)=\frac{c}{\sqrt{\Omega_{\Lambda}}}-{\rm cosn}(\tilde{y}).
\end{equation}
where  $\tilde{y} = \sqrt{|k|}y$ and
\begin{equation}\label{29}
{\rm cosn}(\tilde{y}) = \left\{ \begin{array}{l}
{\rm cos}(y)\,\,\,;~k =  + 1 \\
1\,\,\,\,\,\,\,\,\,\,\,\,\,\,;~~k=0  \\
{\rm cosh}(y)\,\,\,;~k= -1. \\
\end{array} \right.
\end{equation}

\subsection{Non-interacting HDE in CBD model}
Now from conservation equation which is defined in Eq.\;(\ref{16}) and definition of $\rho_{\Lambda}$, Eq.\;(\ref{18}), we can  attain $\omega_{\Lambda}$. So  taking derivative Eq.\;(\ref{18}) w.r.t the time gives
\begin{equation}\label{30}
\dot{\rho}_{\Lambda}=\R_{\Lambda} \left(\frac{\dot{\phi}}{\phi}-2\frac{\dot{L}}{L}\right).
\end{equation}
For calculating (\ref{30}), we must to find out an explicit form for $\phi$, which is the solution of Eq.\;(\ref{14}), but since
  Eq.\;(\ref{14}) is not an independent equation, therefore  obtaining an explicit form for $\p$  is not  possible. So we should remove the extra
  freedome, $\p$,  from the equations of motion.    Recently people have considered the CBD scalar field as a power of the  scale factor which is in a  good agreement with the results of recent observational and experimental data {\citep{40}}. Therefore according to {\citep{40}}, we accept the following  ansatz for  $\phi$ and $f(\phi)$
\begin{equation}\label{31}
\phi=a^{\sigma}, ~~~~~~~f(\phi)=\lambda \phi^{\xi}.
\end{equation}
 In  the BD model one can define $G_{eff}\propto 1/ \p$, where $G_{eff}$ is  the effective Newtonian gravitational constant. Since the observational data requires a constraint on $G_{eff}$ as $|\dot{G}_{eff}/ G_{eff}|\leq 3.32  \times 10^{-20} s^{-1}$  {\citep{41, 41a}}, then  from $|\dot{\p} /\p|=|\dot{G}_{eff}/ G_{eff}| = \s H$, one can restrict the value of $\s$ which  approximately is  $\s \leq 0.01$. On the other hand  there is no any constraint on $\lambda$ and $\xi$, and based on observational evidences  we should find out some constraints on them.
So according to this choice   one can rewrite Eqs.\;(\ref{16}) and (\ref{17}) as
\bea\label{32}
\dot{\rho}_\Lambda &+&\eta H\rho_\Lambda\big(1+\omega_{\Lambda}\big)=0,\\
\dot{\rho}_m&+&\eta H\rho_m\big(1+\omega_m\big)=0,\l{33}
\eea
where $\eta=(3+ \sigma\xi)$. Therefore,  from Eqs.\;(\ref{28}), (\ref{30}) and (\ref{32}) one can obtain $\dot{\rho}_{\Lambda}$ as
\bea\label{34}
\dot{\rho}_{\Lambda}&=&\rho_{\Lambda}H \Big[\sigma-2+\frac{2}{c}\sqrt{\Omega_{\Lambda}}{\rm cosn}(\tilde{y}) \Big],
\eea
 so the EoS parameter in noninteracting case is
\begin{equation}\label{35}
\omega_{\Lambda}=-1-{1\o \eta}{\Big[\sigma-2+\frac{2}{c}\sqrt{\Omega_{\Lambda}}{\rm cosn}(\tilde{y}) \Big]}.
\end{equation}
Note that  in comparison with  EoS in the noninteracting case which  is obtained in \cite{1}, one can see that our results  contain the chameleon effect which appeared in the action.  It is important to note that for  $f(\phi)=1~ (\xi = 0)$ this model reduces to generalized BD model and Eq.\;(\ref{32}) reduces to its respective expression in  BD model.

 From (\ref{35}) it is seen that for $ 0 < c < \sqrt{\Omega_{\Lambda}}{\rm cosn}(\tilde{y})$ and  any arbitrary positive value of $\xi$,
the equation of state parameter, $\omega_{\L}$,  is less than $-1$.  This means that in this case  the EoS parameter of HDE  crosses the phantom divide line $\omega_{\L} = -1$. On the other hand, since $\s\leq 0.01$ and it  is very small than other quantities in Eq.\;(\ref{35}),  then   for $c\geq 1$ the EoS can not crosses the line $\omega_{\L} = -1$ unless $\xi <0$ and $|\xi| > 3/\s$. This means that this model can crosses the phantom divide line with a suitable choices of parameters.

Another useful cosmological parameter is deceleration parameter which  is defined as
\begin{equation}\label{36}
q=\frac{-\ddot{a}}{a H^{2}}=-1-\frac{\dot{H}}{H^{2}}.
\end{equation}
So using Eqs.\;(\ref{10}), (\ref{21}), (\ref{22}), (\ref{25}), (\ref{26}) and (\ref{31}), one can attain $q$ as

\begin{eqnarray}\label{37}
q&=&\frac{1}{(\sigma+2)}\Bigg[ 3f(\phi)\Omega_{\Lambda} \omega_{\Lambda}-\frac{3}{2}\Omega_{V}+(\sigma+1)^{2}\cr
&+&\sigma\Big\{\frac{\omega(\phi) \sigma}{2}-1\Big \}
+\Omega_{k}\Bigg].
\end{eqnarray}

In the present time $f(\p) = \lambda$, $\Omega_{\L} = 0.74$, $\Omega_k = 0.02$, $\s = 0.01$ and we can approximate the function of $\omega(\p) \simeq\omega_0 = 40000$. Then one can get to
\be\l{37'}
q \cong - 1.493\Big[ 0.74|\lambda \omega_{\L}| +{1\over 2}|\Omega_V| -0.41\Big],
\ee
it is obviously seen that, the deceleration parameter can be negative   by a suitable choice of $\lambda$.

In this stage we examine the evolution of $\Omega_{\L}$. Using Eqs.\;(\ref{21}), (\ref{28}) and (\ref{36}) we have
\be\l{37'}
{d\ln(\sqrt{\Omega_{\L}}) \o d\ln(a)}=q + {\sqrt{\Omega_{\L}} \o c}{\rm  cosn}(\tilde{y}).
\ee

Moreover, for checking the evolution of the EoS, $\omega_{\L}$, we have to examine $\dot{\omega}_{\L}$.  We calculate the time derivative of $\omega_{\L}$ and get
\bea\l{372}\nonumber
\dot{\omega}_{\L}&=& -{2H\sqrt{\Omega_{\L} } \o c\eta}\bigg[ \Big\{ q + {1\o c}\sqrt{\Omega_{\L}}{\rm  cosn}(\tilde{y})\Big\}{\rm  cosn}(\tilde{y})\\
 &+&{\sqrt{|k|} \o aH} {\rm sinn(\tilde{y})}\bigg].
\eea
In a spatially flat FLRW Universe, Eq.\;(\ref{372}) becomes
 \bea\l{373}
\dot{\omega}_{\L}= -{2H\sqrt{\Omega_{\L} } \o c\eta}\Big[  q + {1\o c}\sqrt{\Omega_{\L}}\Big].
\eea
Note that in the accelerated expansion phase of the Universe, $q <0$, so Eq.\;(\ref{372}) implies $\dot{\omega}_{\L} <0 $ only for $|q| < \sqrt{\Omega_{\L}}/c$. Therefore in this case the EoS parameter of HDE in  generalized CBD scenario evolves to super- negative value. Indeed this means that if
$|q_1| < \sqrt{\Omega_{\L}}/c$ (where $|q_1|= |q_{\omega_{\L}=-1}|$), then the phase transition take place  from quintessence phase to phantom phase and vice versa
 (i.e., if $|q_1| > \sqrt{\Omega_{\L}}/c$  then the phase transition is take place  from phantom phase to quintessence phase.)
%========================================================================================
%====================Interacting HDE in CBD Cosmology==================================
%========================================================================================
{\subsection{Interacting  HDE in generalized CBD model }}

In this step we consider an interacting between  dark energy candidate (HDE) and (dark) matter. Recently interacting between DE and dark matter attracts very attentions, because the problems which arise from noninteracting case have improved by it. For more conversancy we refer the reader to study the  recent works {\citep{31, 42}} and references there in. In fact the study of this interacting model should be done in a quantum gravity mechanism, but unfortunately such a model is not completely  composed up to now. Following recent researches an interacting term $Q$ has been considered which it play a source-like behavior. Therefore in this case, we can write the modified conservation equations as follows
\bea\label{38}
\dot{\rho}_\Lambda+\eta H\rho_\Lambda\big(1+\omega_{\Lambda}\big)&=&-Q,\\
\dot{\rho}_m+\eta H\rho_m\big(1+ \om_m\big)&=&Q.\l{39}
\eea
Where $Q$  is direct interaction between (dark) matter and DE.
Therefore using Eqs.\;(\ref{34}), (\ref{38}) one can obtain the EoS parameter in the interacting case as
\begin{equation}\label{41}
\omega_{\Lambda}=-1-\frac{1}{\eta}\Big[\sigma-2+\frac{2}{c}\sqrt{\Omega_{\Lambda}}{\rm cosn}(\tilde{y}) +Q/(\R_{\L}H)\Big].
\end{equation}
People have considered some well known candidates for interactive term, which amongst them we consider $Q=3b^{2}H\rho_\L$. Here $b$ is a real constant. So Eq.\;(\ref{41}) becomes
\begin{equation}\label{411}
\omega_{\Lambda}=-1-\frac{1}{\eta}\Big[\sigma-2+\frac{2}{c}\sqrt{\Omega_{\Lambda}}{\rm cosn}(\tilde{y}) +3b^2\Big].
\end{equation}
By comparing Eq.\;(\ref{411}) with Eq.\;(\ref{35}), one  can see that the direct interaction between (dark) matter and DE helps phantom divide line crossing in this model.  Whereas the EoS parameter in this case differs from EoS in noninteracting case we should obtain deceleration parameter in this case. Therefore from Eqs.\;(\ref{24}), (\ref{29}) one can attain

\begin{eqnarray}\label{43}\nonumber
q&=&\frac{1}{(\sigma+2)}\Bigg[ 3f(\phi)\Omega_{\Lambda} \omega_{\Lambda}-\frac{3}{2}\Omega_{V}+(\sigma+1)^{2}\\
&+&\sigma(\frac{\omega(\phi) \sigma}{2}-1)+\Omega_{k}\Bigg].
\end{eqnarray}

Notice that the  deceleration parameter in interacting case explicitly contain the  CBD scalar field function as the same as  noninteracting case. We should emphasize that  the $q$ which obtained in noninteracting case in the  {\citep{1}} is not a function of the $f(\phi)$ and is mentioned as a minor mistake. Note that the form of the deceleration parameter in interacting case is similar to non-interacting case but the equations of state parameter against the EoS. Also  the evolution equation of dimensionless parameter, $\Omega_{\L}$, is such as  (\ref{37'}) versus new $q$, (\ref{43}),  then it is not necessary to find them again.

\section{Conformal-age-like length as an IR cut off}
In this section we take a conformal age like parameter as the characteristic
length scale $L$, which is defined in the flat FLRW Universe as follows
\be\l{44}
L= {1\o a^4(t)}\int_0^t dt' a^3(t').
\ee
where $a(t)$ is the scale factor {\citep{43}}. So taking derivative with respect to the cosmic time from $L$ and using  Eqs.\;(\ref{18}) and  (\ref{27}) we find
\be\l{45}
\dot{L} = -4LH + {1\o a},
\ee
\subsection{Non-interacting HDE in CBD model}
Now from conservation equation, Eq.\;(\ref{16}), and definition of $\rho_{\Lambda}$,  Eq.\;(\ref{18}), we can  attain $\omega_{\Lambda}$.
\be\l{46}
\omega_{\Lambda}=-1-{1\o \eta}{\Big[\sigma+8-\frac{2}{ca(t)}\sqrt{\Omega_{\Lambda}} \Big]}.
\ee
It is seen that at   present time and for $c\geq 1$, $\omega_{\L} <-1$ i.e., this model can cross the phantom divide line. Also  the deceleration parameter is

\begin{eqnarray}\label{47}
q&=&\frac{1}{(\sigma+2)}\Bigg[ 3f(\phi)\Omega_{\Lambda}\Big\{ -1-{1\o \nonumber \eta}{\big(\sigma+8-\frac{2}{ca(t)}\sqrt{\Omega_{\Lambda}} \big)}\Big\}\\ &-&\frac{3}{2}\Omega_{V}+(\sigma+1)^{2}+\sigma(\frac{\omega(\phi) \sigma}{2}-1)\Bigg].
\end{eqnarray}

In order to gain better insight we focus on the situation which $ \omega(\p) = \omega_0 = 40000$. So in this case
 one can obtain the equation of motion for dimensionless parameter of dark energy density as
\bea\l{48}\nonumber
\theta\Om'_{\L}&=&\Om_{\L}\Bigg[\Big\{ \eta(1+\om_m)+\s +8 -{2\o ac}\sqrt{\Om_{\L}}\Big\}(\theta - \Om_{\L})\\
 &+& {1\o 2}(\xi+\nu)\s\Om_V\Bigg],
\eea
where $\theta =1-\s(\s \om_0/6 -1)$.
Accepting that the Universe had the inflation epoch  with $\om_m=-1$, the radiation epoch  with $\om_m=\om_r =1/3$ , and the matter dominant epoch  with $\om_m =0$ before entering into the accelerating expansion phase, we want have an approximately investigation  of $L$ and $\Om_{\Lambda}$ in these three epochs. In fact we want to obtain the fractional density of dark energy in the early universe from directly calculation.
In this part of our work we assume the EoS parameter, $\om_m$ is constant in all epochs. So according to Eq.\;(\ref{33}) we have $\R_m = \R_0 a^{-\eta(1+\om_m)}$  and from Eq.\;(\ref{9}) we have
\be\l{49}
H^2\simeq \big({\lambda \R_0 \o 3\theta}\big) a^{-2\zeta} + \big(\frac{M^5}{6\theta}\big) a^{-2\sigma},
\ee
where the density of dark energy is ignored in early time. And since
\be\l{50}
\zeta= {1\o 2}(3 + \sigma + \eta\om_m) \gg \sigma,
\ee
then the second term   in the right hand side of Eq.\;(\ref{49}) is negligible with respect to the first one and this relation reduced to
\be\l{51}
H^2\approx \big({\lambda \R_0 \o 3\theta}\big) a^{-2\zeta}.
\ee
Therefore, from (\ref{44}) we have
\be\l{52}
L = \big({a_i\o a}\big)^4 L_i + {1 \o 3+\zeta} \Big[ {1\o Ha} - {1\o H_ia_i}\big({a_i\o a}\big)^4
\Big],
\ee
where $L_i = (1/a_i^4)\int_0^{t_i} dt a^3(t)$ and  subscript "$i$" indicates the value of corresponding quantity at the beginning of the epoch. Also
$a$  is the scale factor out of matter dominant epoch. This means   $(a_i/a)^4\ll1$, so we have
\be\l{53}
L\sim \big( {1 \o 3 + \zeta}\big) {1 \o Ha},
\ee
and from (\ref{21}) we have
\be\l{54}
\Om_{\L} \sim \lambda(3 + \zeta)^2 c^2 a^{(2+\s\xi)},
\ee
Since $c$ is  at order of unity and $\s$ is very small, then Eq.\;(\ref{54}) shows that the value of dimensionless parameter of dark energy before the accelerating expansion phase is very small. In fact Eq.\;(\ref{54}) is an approximated solution for early time until matter dominant, so we can not match  it with the present value of fractional parameter of dark energy. Therefore to obtain the reasonable  value of parameters, we should find the exact  solution of Eq.\;(\ref{48}) and matching it with the present value of dark energy.
\subsection{Interacting HDE with conformal age like length }
  In this section we consider an interaction between DE and the  matter  similar to Subsection.B in pervious  Section.
  In this case the modified conservation equations  are Eqs.\;(\ref{38}) and (\ref{39}). Then using Eqs.\;(\ref{30}), (\ref{38}) and (\ref{45})
    one  obtains  the EoS parameter as
\begin{equation}\label{57}
\omega_{\Lambda}=-1-{1\o \eta}{\Big[\sigma+8-\frac{2}{ca(t)}\sqrt{\Omega_{\Lambda}}+Q/ \R_\L H \Big]},
\end{equation}
and for  $Q=3b^{2}H\rho_\L$,  Eq.\;(\ref{57}) becomes
\begin{equation}\label{57'}
\omega_{\Lambda}=-1-{1\o \eta}{\Big[\sigma+8-\frac{2}{ca(t)}\sqrt{\Omega_{\Lambda}}+3b^2 \Big]}.
\end{equation}
 Eq.\;(\ref{57'}) shows  that the direct interaction between (dark) matter and dark energy helps to crossing the phantom divide line.

Finally,  the deceleration parameter of HDE model in CBD scenario for conformal age like length  is  attained as

\begin{eqnarray}\label{58}
q&=&\frac{1}{(\sigma+2)}\Bigg[ 3f(\phi)\Omega_{\Lambda}\Big\{ -1-{1\o \nonumber \eta}{\big(\sigma+8-\frac{2}{ca(t)}\sqrt{\Omega_{\Lambda}} \big)}\\
&+&Q/ \R_\L H \Big\} -\frac{3}{2}\Omega_{V}+(\sigma+1)^{2}+\sigma(\frac{\omega_0 \sigma}{2}-1)\Bigg].
\end{eqnarray}

In the present time $f(\p) = \lambda$, $\Omega_{\L} = 0.74$, $\Omega_k = 0.02$, $\s = 0.01$ and we can approximate the function of $\omega(\p) \simeq\omega_0 = 40000$. Then one can get to
\be\l{59}
q \cong - 1.493\Big[ 0.74|\lambda \omega_{\L}| +{1\over 2}|\Omega_V| -0.41\Big],
\ee
it is obviously seen that, the deceleration parameter can be negative   by a suitable choice for $\lambda$. The differences between Eq.\;(\ref{59}) and Eq.\;(\ref{37'}) is  $|\om_{\L}|$ in which the value  of $|\om_{\L}|$ in   Eq.\;(\ref{59}) is bigger than it in Eq.\;(\ref{37'}).

%====================================================================
%======================= conclusion===================================
%=====================================================================

{\section{Conclusion and discussion}}

In this paper, we have considered holographic dark energy (HDE) in the Brans-Dicke cosmological model with a non-minimal coupling between the chameleon scalar field and matter. This model has a dynamical time dependent scalar field which behave like dark energy and then it might produce cosmic acceleration. However, we suppose the matter sector consists of matter and dark energy, so this means that we have assumed an interaction between the scalar field and dark energy too. This fact is clearly seen in Eqs.\;(\ref{9}) and (\ref{10}). Actually, non-minimal coupling between the scalar field and matter field generalizes the conservation equation of energy density which is diagnosed wrong in {\citep{1}}.

We studied the interacting and non-interacting cases of this model for two different infrared cutoffs; future event horizon and conformal-age-like length. We obtained the EoS parameter, deceleration parameter and fraction parameter of dark energy for two mentioned cutoffs and we found that phantom crossing is possible in both of cutoffs by tuning the free parameters of the model. Note that in almost every cosmological model the fine tuning of parameters is necessary and our model  also is not exception. At the end, comparing the obtained results for two mentioned cutoffs, show that the phantom crossing with conformal-age-like length is more possible than future event horizon cutoff.

\begin{acknowledgment}
%\acknowledgment

The authors would like to thank  Ali Aghamohammadi for his useful discussions.

\end{acknowledgment}

%%%%%%%%%%%%%%%%%%%%%%%%%%%%%%%%%%%%%%%
%%%%%%%%%%%%%%%%%%%%%%%%%%%%%%%%%%%%%%%%%
%%%%%%%%%%%%%%%%%%%%%%%%%%%%%%%%%%%%%%%%%%
%\nocite{*}
%\bibliographystyle{spr-mp-nameyear-cnd}
%\bibliography{myref}
%\bibliography{biblio-u1}

\

\end{document}